%% file: main.tex
\documentclass[envcountsect,envcountsame,runningheads,a4paper]{llncs}
\usepackage{FM2021}

\begin{document}

\title{Hybrid Systems Verification with Isabelle/HOL:\\
  Simpler Syntax, Better Models, Faster Proofs}
\titlerunning{Hybrid Systems Verification with Isabelle/HOL}
\author{Simon Foster\inst{1} \and Jonathan Juli\'an Huerta y Munive\inst{2} \and \\ Mario
  Gleirscher\inst{3} \and  Georg Struth\inst{4}} 
\authorrunning{Foster, Huerta y Munive, Gleirscher, Struth}
\institute{University of York, UK \and University of Copenhagen, Denmark \and University of Bremen, Germany \and University of Sheffield, UK}

\maketitle

\begin{abstract}
  We extend a semantic verification framework for hybrid systems with
  the Isabelle/HOL proof assistant by an algebraic model for hybrid
  program stores, a shallow expression model for hybrid programs and
  their correctness specifications, and domain-specific deductive and
  calculational support. The new store model yields clean separations
  and dynamic local views of variables, e.g. discrete/continuous,
  mutable/im\-mutable, program/logical, and enhanced ways of
  manipulating them using combinators, projections and framing. This
  leads to more local inference rules, procedures and tactics for
  reasoning with invariant sets, certifying solutions of hybrid
  specifications or calculating derivatives with increased proof
  automation and scalability. The new expression model provides more
  user-friendly syntax, better control of name spaces and interfaces
  connecting the framework with real-world modelling languages.
\end{abstract}

\section{Introduction}
\label{sec:intro}
\input{sec/intro}

\section{Semantic Preliminaries}
\label{sec:prelim}
\input{sec/prelim}

\section{Hybrid Store Components}

\label{sec:hyb-store}
\input{sec/hyb-store}

\section{Shallow Expressions Component \hfill\isalink{https://github.com/isabelle-utp/Shallow-Expressions}} 

\label{sec:prog-exp}

\input{sec/prog-exp}

\section{Dynamical Systems Components}
\label{sec:dyn-sys}

\input{sec/diff-eq}

\section{Reasoning Components}
\label{sec:proof}

\input{sec/proof}

\section{Examples}
\label{sec:case-studies}
\input{sec/case-studies}

\section{Related Work, Conclusions, and Future Work}
\label{sec:concl}

\input{sec/concl}

\section*{Acknowledgements}

This work is funded by UKRI-EPSRC project
\href{https://www.cs.york.ac.uk/circus/CyPhyAssure/}{CyPhyAssure} (grant reference EP/S001190/1);
Novo Nordisk Fonden Start Package Grant~(NNF20OC0063462); and the
\href{https://www.york.ac.uk/assuring-autonomy/}{Assuring Autonomy International Programme}~(AAIP;
grant CSI:Cobot), a partnership between Lloyd's Register Foundation and the University of York.

\bibliographystyle{splncs}
\bibliography{main.bib}

\end{document}

%% file: sec/intro.tex
Deductive verification of hybrid systems with interactive proof assistants like
Coq, HOL or Isabelle is currently gaining
traction~\cite{RickettsMAGL15,WangZZ15,MuniveS18,MuniveS19,Foster19b-HybridRelations,FosterMS20,afp:hybrid,afp:linear}. Such
tools have reached a level of maturity, proof power, and mathematical library
support that makes the development of formal methods fast, dependable and
competitive.  With Isabelle/HOL, an impressive theory stack for ordinary
differential equations (ODEs)~\cite{ImmlerT19} has been combined with algebras
of programs and concrete hybrid program semantics into a semantic framework for
reasoning about hybrid systems~\cite{MuniveS19}. Its development has so far
focused on mathematical groundwork, exploring Isabelle's mathematical and theory
engineering facilities, supporting verification workflows for hybrid systems,
and implementing a tool prototype.

Here we report on its transformation into a formal method: (1) We
supply a better hybrid program store model to reason about discrete
and continuous, mutable and immutable, logical and program variables
more succinctly (\S\ref{sec:hyb-store}). It supports local reasoning
about parts of stores through projection and framing techniques, and
generation of fresh variables~\cite{Foster2020-LocalVars}. (2) Using
this model, we derive deductive and calculational rules for hybrid
system verification, including inference rules \`a la differential
dynamic logic (\dL)~\cite{Platzer18} (\S\ref{sec:dyn-sys}).  Rules are
enhanced with framing conditions for local reasoning about mutable
continuous variables. We supply a new ghost rule for invariant sets, a
new frame rule \`a la separation logic, and more effective tactics,
for instance for calculating framed Fr\'echet derivatives
(\S\ref{sec:proof}). (3) Using Isabelle's syntax translation
mechanisms we create a simpler modelling and specification language
within the shallow embedding of our framework
(\S\ref{sec:prog-exp}). It is meant to be extensible to modelling
languages such as Modelica and computer algebra systems such as
Mathematica. These contributions are explained through examples
(\S\ref{sec:case-studies}); additional ones are highlighted throughout
the article. Next we outline the main features of the framework to
contextualise these contributions and prepare for the technical
sections.

The framework~\cite{MuniveS19} has been formalised as a shallow
embedding using Isabelle's own language and types. The benefits of
shallowness are well
documented~\cite{BoultonGGHHT1992,WildmoserN04}. A drawback addressed
by our expression model is that syntactic properties, for instance of
mutable variables, may be hard to capture. The framework has three
semantic layers that can be modified compositionally: abstract
algebras of programs are used for deriving structural program
transformation laws and verification condition generation
(VCG)~\cite{NipkowK14} by equational reasoning.  Isabelle's type
polymorphism allows their instantiation, for example with state
transformer semantics and further with concrete semantics of hybrid
program stores---the level at which basic commands such as assignments
are modelled.  The store extension to hybrid programs is obtained by a
state transformer semantics for basic (continuous) evolution
commands. As in \dL, these commands specify ODEs (via vector fields)
and guards imposing boundary conditions on their state spaces. The
associated state transformer then maps initial states of evolutions to
their guarded orbits, or more general sets of reachable states. Hybrid
stores were modelled so far~\cite{MuniveS19} as real-valued functions
from program variables, using $\mathbb{R}^n$ with a fixed finite set
of natural numbers as variable names for technical reasons. Our new
hybrid store model supports more general name spaces, dynamic stores
and other implementations via records or monads.

Users of the framework need to specify hybrid programs and standard correctness
specifications using pre/postconditions and loop invariants. Two main workflows
are then supported: (1) If ODEs have unique solutions, one can certify them or
rely on automatic certifications in special cases~\cite{Munive20,afp:linear},
then trigger automatic VCG and discharge the remaining verification conditions
(VCs) by reasoning about solutions in state spaces. (2) More generally, one can
assert invariant sets~\cite{Teschl12} for ODEs, trigger VCG and then reason
abstractly about these invariants to discharge the VCs. So far, these two
workflows have relied on Isabelle's internal support for equational reasoning,
which seems natural for mathematicians or engineers. Our new components make
them more automatic by reasoning locally with certification conditions,
invariant assertions or derivatives, and more structured by supporting
data-level reasoning with invariants using the derived \dL-style inference
rules, for those who prefer this approach. Capturing fresh variables is
instrumental for our new \dL-inspired ghost rule; projecting on mutable
variables simplifies proof obligations and localises reasoning about such
variables as does separation logic, but by different means.

The general framework has been tested successfully on a large set of hybrid
verification benchmarks~\cite{MitschMJZWZ20}; our new components, integrated
into an accessible simplified version of the framework, can be found
online\footnote{\href{https://github.com/isabelle-utp/Hybrid-Verification}{\texttt{github.com/isabelle-utp/Hybrid-Verification}},
also clicking our \isalogo\, icons.}.

%% file: sec/prelim.tex
We first recall the basics of state and predicate transformers, the
semantics of evolution commands and the set-up for our new hybrid
store model~\cite{MuniveS19,Foster2020-IsabelleUTP}. The semantics can
be motivated using the hybrid program syntax of \dL~\cite{Platzer18},
$X\, ::=\, x:=e \mid x' = f \, \&\, G \mid ?P\mid X\mathrel{;} X\mid
X+X\mid X^*$, which, beyond standard constructs of dynamic logic,
features an \emph{evolution command} $x' = f \, \&\, G$. It specifies
a vector field $f:T\rightarrow \src\rightarrow \src$ with time domain
$T$ over a state space $\src$ and a \emph{guard}
$G:\src\rightarrow \bool$, a predicate modelling boundary conditions.

\vspace{-\baselineskip}

\subsubsection{State and predicate transformers.}

We model programs as \emph{state transformers}
$\alpha:\src\rightarrow \power \src$, arrows of the Kleisli category of the
powerset monad. (Forward) Kleisli composition
$(\alpha\circ_K \beta)\, x = \bigcup\{\beta\, y\mid y \in \alpha\, x \}$ models
sequential composition, the program skip is the monadic unit
$\eta_\src\, x = \{x\}$, abort is $\lambda x.\, \emptyset$,
nondeterministic choice $\cup$ on functions and finite iteration
$\alpha^{\ast}\, x = \bigcup_{i\in\mathbb{N}} \alpha^{i}\, x$, with powers
defined using $\circ_K$. Tests (and assertions) are subidentities
(functions $P\le \eta_\src$, with $\le$ extended pointwise) mapping
any $x\in\src$ either to $\{x\}$ or $\emptyset$.  They are isomorphic
to sets and predicates. Deterministic functions $\src\rightarrow \src$
are embedded into $\src\rightarrow \power \src$ via
$\assigns{\alpha} = \eta_\src \circ \alpha$. Backward diamond operators
(disjunctive predicate transformers with contravariant composition)
are Kleisli extensions of state transformers, forward box or $\wlp$
operators are their right adjoints on the boolean algebra of tests:
$\wlp\ \alpha\, Q =\{x \mid \alpha\, x \subseteq Q\}$ for any program $\alpha$ and
test $Q$. We often write partial correctness assertions
$P\subseteq \wlp\, \alpha\, Q$ as $\{P\}\, \alpha\, \{Q\}$.

The laws of propositional Hoare logic (ignoring assignments) are
derivable in this semantics. Nevertheless, VCG with
$\wlp\, (\alpha\circ_K \beta) = \wlp\, \alpha\circ \wlp\, \beta$ and
$\wlp\, (\mathsf{if}\ P\ \mathsf{then}\, \alpha\, \mathsf{else}\,
\beta)\, Q = (\overline{P}\cup \wlp\, \alpha\, Q)\cap (P\cap \wlp\,
\beta\, Q)$ is more effective, while the standard Hoare rule can be
used for loops decorated with invariants.

\vspace{-\baselineskip}

\subsubsection{Continuous dynamics.}

The evolution of continuous systems~\cite{Teschl12} is often modelled
by \emph{(local) flows} $\varphi: T\rightarrow \src\rightarrow \src$,
where $T\subseteq \real$ models time and $\src$ a state space. Flows
are assumed to be $C^1$-functions, and monoid actions if $T=\real$:
$\varphi (t+t') = \varphi\, t \circ \varphi\, t'$ and
$\varphi\, 0 = \mathit{id}_\src$.  A \emph{trajectory}
$\varphi_s: T\rightarrow \src$ of $\varphi$ at $s\in \src$ is then a
curve $\varphi_s\, t= \varphi\, t\, s$, and its \emph{orbit} at
$s\in \src$ is given by state transformer
$\gamma^\varphi: \src\rightarrow \power \src, s \mapsto \power\,
\varphi_s\, T$, which maps any state $s$ to the set of states on the
trajectory passing through it. Flows are typically solutions to
\emph{initial value problems} for systems of ODEs, which specify a
\emph{vector field} $f:T\rightarrow \src\rightarrow \src$ assigning
vectors to points in space-time, and an initial value $s\in \src$ at
$t_0\in T$. A \emph{solution} is then a $C^1$-function $X$ that
satisfies $X'\, t= f\, t\, (X\, t)$ and $X\, t_0=s$.  For $f$
continuous, existence of $X$ is guaranteed by the Peano theorem. Yet
$f$ must be Lipschitz continuous to guarantee uniqueness via the
Picard-Lindel\"of theorem, which provides intervals $U\, s\subseteq T$
where solutions exist around $t_0$ for each $s\in \src$ and gives rise
to flows when $t_0=0$.

\vspace{-\baselineskip}

\subsubsection{Semantics of evolution commands.}

Orbit maps $\gamma^\varphi:\src\rightarrow \power \src$ are state
transformers, but we generalise to include guards and to continuous
vector fields.  Our state transformer for evolution commands maps each
state $s\in \src$ to the set of reachable states (or \emph{generalised
  orbit})
\begin{equation*}
  (x' = f \, \&\, G)\, s = \{X\, t\mid t\in U\, s\land 
  (\forall\tau\in\downarrow_{U\, s} t.\ G (X\, \tau))\land
  X'\, t= f\, t\, (X\, t)\land X\, t_0=s\},
  \end{equation*}
  where $\mathop{\downarrow_{U\, s}} t$ is the down-closure of $t$ in
  $U\, s$, such that $t_0\in U\, s$. It constrains the domain of
  existence of the solutions $X$. If we know the flow $\varphi$ for
  $f$, then this semantics reduces to
  $(x' = f \, \&\, G)\, s = \{\varphi\, t\, s\mid t\in U\, s\land
  (\forall\tau\in\mathop{\downarrow_{U\, s}} t.\ G (\varphi\, \tau\,
  s)) \}$.
  
  This state transformer maps any $s\in\src$ to all reachable states
  along any solution to $f$. Thus $\wlp\ (x' = f \, \&\, G)\, Q$ holds
  iff for every $t\in U\, s$, if $G\, (X\, \tau)$ holds for all
  $\tau\in\mathop{\downarrow_{U\, s}} t$, then $Q\, (X\, t)$ holds as
  well. Computing $\wlp$'s within the remits of Picard-Lindel\"of is
  then straightforward: users only need to supply and certify the flow
  $\varphi$ within the first workflow outlined in \S\ref{sec:intro}.

  \vspace{-\baselineskip}

\subsubsection{Invariant sets.}
  
Instead of analytic solutions, one can use the generalised orbit
semantics in combination with generalised invariant sets in the second
workflow. Invariant sets~\cite{Teschl12} are preserved by the orbit
map of the dynamical system. With guards, more generally,
$I\subseteq \src$ is an \emph{invariant set} of $(x' = f \, \&\, G)$
whenever $(x' = f \, \&\, G)^\dagger\, I \subseteq I$, where
$(-)^\dagger$ indicates Kleisli extension and yields a backward modal
diamond operator. The adjunction mentioned translates this property
into $I\subseteq \wlp\, (x' = f \, \&\, G)\, I$, the standard format
for invariance reasoning with predicate
transformers~\cite{MuniveS19}. Intuitively, such invariants
characterise regions of the state space that contain all orbits that
respect $G$ and have a point inside them. It then suffices to assert
suitable invariants in order to verify a correctness specification.
We discuss invariance techniques for evolution commands in
\S\ref{sec:dyn-sys}.

 \vspace{-\baselineskip}

\subsubsection{Lenses.}

\newcommand{\lens}{\lambda}

Our algebraic model for hybrid stores is based on
lenses~\cite{Foster2020-IsabelleUTP,Optics-AFP}, a tool for
manipulating program stores or state spaces, which comes in many
variants and guises~\cite{Oles82,BackW98,Foster09}, in support of
algebraic reasoning about program
variables~\cite{FosterZW16,Foster2020-IsabelleUTP} and local reasoning
about store
shapes~\cite{Foster2020-LocalVars}. \hfill\isalink{https://github.com/isabelle-utp/Optics/blob/main/Lens_Laws.thy}

A \emph{lens} $\lens : \view \lto \src$ is a pair
$(\lget_\lens : \src \rightarrow \view, \lput_\lens : \view
\rightarrow \src \rightarrow \src)$ such that
\begin{equation*}
  \lget_\lens~(\lput_\lens~v~s) = v, \qquad \lput_\lens~v \circ
\lput_\lens~v' = \lput_\lens~v, \qquad
\lput_\lens~(\lget_\lens s)~s = s,
\end{equation*}
for all $v\in \view$ and $s\in \src$. Lenses are arrows in a category,
as explained in \S\ref{sec:hyb-store} below. They admit many
interpretations. Typically, $\src$ is a set of program stores. Yet
$\view$ could be a smaller or simpler set of stores, $\lget\, s$ could
forget part of $s\in \src$ and $\lput\, v\, s$ overwrite part of $s$
with $v\in\view$. Otherwise, $\view$ could be a value domain,
$\lget\, s$ could look up a value $v$ in $s$ and $\lput\, v\, s$ could
update $s$ with $v$.

We use \emph{variable lenses} $x:\view \lto \src$ to model program variables
$x$, $\lget_x\, s$ to look up values in $\view$ in stores $s$ and
$\lput_x\, v\, s$ to update values in $s$ by $v$. They can be implemented
differently as concrete stores~\cite{FosterZW16}. A \emph{variable assignment}
$x:=e$, with expression $e$ represented semantically as a function
$\src \rightarrow \view$, is then a state transformer
$(x:=e) = \assigns{\lambda s.\, \lput_x\, (e\, s)\, s}$, and
$\wlp\, (x:=e)\, Q = \lambda s.\, Q\, (\lput_x\, (e\, s)\, s)$. This suffices
for VCG for hybrid programs without evolution commands.

%% file: sec/hyb-store.tex
Lenses enhance our hybrid store models. They provide a unifying
interface for heterogeneous variables, allow modelling frames, i.e.,
sets of mutable variables, and projecting parts of the global state to
vector spaces of continuous evolutions. These projections are based on
three lens combinators: composition, sum and quotient. This makes
reasoning about hybrid stores local.

\vspace{-\baselineskip}

\subsubsection{States and variables.}

Many modelling and programming languages support modules with local
variable and constant declarations. We have implemented an Isabelle
command that automates the creation of hybrid stores. \hfill\isalink{https://github.com/isabelle-utp/Optics/blob/main/Dataspace_Example.thy}

{\small
\vspace{-1ex}
\begin{alltt}
\isakwmaj{dataspace} sys =
  \isakwmin{constants} c\(\sb{1}\)::C\(\sb{1}\)..c\(\sb{n}\)::C\(\sb{n}\) \isakwmin{assumes} a\(\sb{1}\):P\(\sb{1}\)..a\(\sb{n}\):P\(\sb{n}\) \isakwmin{variables} x\(\sb{1}\)::T\(\sb{1}\)..x\(\sb{n}\)::T\(\sb{n}\)
\end{alltt}}
\vspace{-1ex}
\noindent It has constants $c_i\!:\!C_i$, named constraints
$a_1\!:\!P_i$ and state variables $x_i\!:\!T_i$. Inside, we can create
local definitions, theorems and proofs, which are hidden, but
accessible using its namespace. Internally, a \isakwmaj{locale} with
fixed constants and assumptions is created. Each variable is a lens
$x_i : T_i \lto \src$, using abstract store $\src$ with the lens
axioms as locale assumptions. We also generate independence
assumptions~\cite{Foster2020-IsabelleUTP} that distinguish different
variables semantically. Lenses $\lens,\lens':\view \lto \src$ are
\emph{independent}, $\lens \lindep \lens'$, if
$\lput_\lens~u \circ \lput_{\lens'}~v = \lput_{\lens'}~v \circ
\lput_\lens~u$, for all $u,v\in \view$, that is, their actions commute
on all states.

\vspace{-\baselineskip}
\subsubsection{Substitutions.}
We obtain cleaner program specifications with the notation
$\sigma(x\!\smapsto\!e) = \lambda s.\,\lput_x\, (e\, s)\, (\sigma\,s)$. With
this, we can describe assignments as sequences of updates: for variable lenses
$x_i : \view_i \lto \src$ and ``expressions'' $e_i : \src \rightarrow \view_i$,
$[x_1 \leadsto e_1, x_2 \leadsto e_2, \cdots] = id(x_1 \leadsto e_1)(x_2
\leadsto e_2)\cdots$. Implicitly, any variable $y$ not mentioned in such a
semantic ``\emph{substitution}'' $\sigma:\src\to\src$ is left unchanged:
$y \smapsto y$. We write e.g. $e[v/x] = e \circ [x \smapsto v]$ for the
application of substitutions to expressions. These notations unclutter program
specifications significantly, e.g. $(x := e) = \assigns{[x \leadsto e]}$ and
$\wlp~\assigns{[x\leadsto e]}~Q = Q[e/x]$. Crucially, the Isabelle simplifier
can reorder and reduce substitutions to support syntactic manipulation of
variables during VCG~\cite{Foster2020-IsabelleUTP}. We can extract assignments
for $x$ with $\substlk{\sigma}\,x = \lget_x \circ \sigma$, and so e.g.
$\substlk{[x \smapsto e_1, y \smapsto e_2]}\, x$ reduces to $e_1$ when
$x \lindep
y$. \isalink{https://github.com/isabelle-utp/Shallow-Expressions/blob/main/Substitutions.thy}

\vspace{-\baselineskip}

\subsubsection{Vectors and matrices.} The lens category admits a
forward \emph{lens composition}
$\lens_1 \lcomp \lens_2:\src_1\lto \src_3$, for
$\lens_1:\src_1 \lto \src_2$, $\lens_2:\src_2\lto\src_3$ and
units $\lone_\src : \src \lto
\src$~\cite{Foster09,Foster2020-IsabelleUTP}.  We do not
reproduce the formal definition. Intuitively, $\lcomp$ selects a part
of a larger store shape, as we will shortly demonstrate.

Vectors and matrices are supported by \textsf{HOL-Analysis}. We supply
notation \texttt{[[x11,...,x1n],...,[xm1,...,xmn]]} for matrices and
means for accessing coordinates of vectors via hybrid program
variables~\cite{FosterGC20}. We view vectors in $\real^n$ as part of larger hybrid
stores, hence as lenses $\real^n\lto \src$, and project onto
coordinate $v_k$ of any vector $\vec{v}$ in $\real^n$ using lens
composition and a \emph{vector lens} for $v_k$. \hfill\isalink{https://github.com/isabelle-utp/Hybrid-Library/blob/6b37e352181fb5613cc9a960df6aed12d68cf370/Cont_Lens.thy\#L117}
\begin{equation*}
  \Pi(k : [n]) = ((\lambda s.\, \textit{vec-nth}~s~k) : A^n \rightarrow A,
(\lambda v~s.\, \textit{vec-upd}~s~k~v) : A \rightarrow A^n \rightarrow A^n),
\end{equation*}
where $[n]=\{1..n\}$, $A^n \cong [n] \rightarrow A$, and the lookup
function $\textit{vec-nth}$ and update function $\textit{vec-upd}$
come from \textsf{HOL-Analysis}. Then, for example,
$v_x = \Pi(1) \lcomp \vec{v}$ and $v_y = \Pi(2) \lcomp \vec{v}$ for
$\vec{v} : \real^2 \lto \src$, using $\lcomp$ to first select the
variable $\vec{v}$ and then the vector-part of the hybrid store.
Obviously, $\Pi(i) \lindep \Pi(j)$ iff $i \neq j$.

We can specify ODEs and flows via substitutions:
$[\vec{p}\!\leadsto\!\vec{v}, \vec{v}\!\leadsto\!\vec{a},
\vec{a}\!\leadsto\!0]$, e.g., specifies a vector field for lenses $\vec{p}$,
$\vec{v}$, $\vec{a} : \real^2 \lto \src$. Though ostensibly syntactic objects,
these substitutions are semantically functions $\src \to \src$, and consequently
can be used with Isabelle's ODE components~\cite{Immler12,HolzlIH13}.

\vspace{-\baselineskip}

\subsubsection{Frames.}

Lenses support algebraic manipulations of variable frames. A frame is
a set of variables that may be mutated by a program within a given
context. We first show how variable sets can be modelled via lens
sums. Then we define a predicate that characterises the immutable
variables for programs~\cite{Foster2021-IsaSACM}. Equipped with this
we derive a frame rule in the style of separation logic that embodies
local reasoning with framed variables.

Variable lenses can be combined into lenses for variable sets with \emph{lens
  sum}~\cite{Foster2020-IsabelleUTP}
$\lens_1 \lplus \lens_2: \view_1\times \view_2 \lto \src = (\lambda (s_1,
s_2).\, (\lget_{\lens_1}, \lget_{\lens_2}), \lput_{\lens_1} \circ
\lput_{\lens_2})$, which is defined for $\lens_1 : \view_1\lto \src$,
$\lens_2 :\view_2 \lto \src$ if $\lens_1 \lindep \lens_2$. This combines two
independent lenses, and can model composite variables, e.g.
$(x, y) = x \lplus y$, which can be decomposed by the simplifier:
$[(x, y)\!\smapsto\!(e_1, e_2)] = [x\!\smapsto\!e_1, y\!\smapsto\!e_2]$. We can
also use it to specify finite sets: $\{x, y, z\}$ as $x \oplus (y \oplus z)$,
yet each each variable in the sum may have a different type, e.g.
$\{v_x, \vec{p}\}$ is a valid and well-typed construction. \hfill\isalink{https://github.com/isabelle-utp/Optics/blob/6e24cde61989a79f7601acc537dd2ee9fdf3f4f6/Lens_Algebra.thy\#L38}

Sums of lenses cannot be directly related with sets: they are only associative
and commutative up-to isomorphism of cartesian products. We thus define a
\emph{lens preorder}~\cite{Foster2020-IsabelleUTP},
$\lens_1 \preceq \lens_2 \Leftrightarrow \exists \lens_3.\, \lens_1 = \lens_3
\lcomp \lens_2$ that captures the part-of relation between
$\lens_1 : \view_1 \lto \src$ and $\lens_2 : \view_2 \lto \src$, e.g.
$v_x \le \vec{v}$ and $\vec{p} \le \vec{p} \lplus \vec{v}$.  \emph{Lens
  equivalence} ${\lequiv} = {\preceq}\cap {\succeq }$ then identifies lenses
that have the same shape in the store. Then, for variable set lenses up-to
$\lequiv$, $\lplus$ models $\cup$, $\lindep$ models $\notin$, and $\preceq$ models
$\subseteq$ or $\in$, which we can use to construct variable frames. \hfill\isalink{https://github.com/isabelle-utp/Optics/blob/main/Lens_Order.thy}

Let $A:\view\lto\src$ be a lens modelling a variable set.
For $s_1,s_2\in \src$ let $s_1\approx_A s_2$ hold if $s_1=s_2$ up-to the values
of variables in $A$. For $\alpha:\src\Rightarrow \power \src$ define
$\alpha\nmods A \Leftrightarrow \forall s_1\in\src\, s_2\in \alpha\, s_1
\Rightarrow s_1\approx_A s_2$, i.e., the mutable variables in $\alpha$ are not
in $A$. Then $ (x := e) \nmods A$ whenever $x \lindep A$ and, recursively,
$(\alpha \relsemi \beta) \nmods A$ and
$ (\mathsf{if}~P~\mathsf{then}~\alpha~\mathsf{else}~\beta) \nmods A$ when
$\alpha\nmods A$ and $\beta \nmods A$. Also, $A \le B$ and $\alpha \nmods B$
implies that $\alpha \nmods A$. \hfill\isalink{https://github.com/isabelle-utp/Hybrid-Verification/blob/7baedd092dff0182336ef0bb6251fd8beff6a1cc/HS_Lens_Spartan.thy\#L381}

Similarly, we use lenses to describe when a variable does not occur freely
in an expression or predicate as
$A \unrest e \Leftrightarrow \forall v.\, e~(\lput_A~v) =
e$~\cite{Foster2020-IsabelleUTP}. We can now derive
a variant of separation logic's frame rule, which epitomises local
reasoning: \hfill\isalink{https://github.com/isabelle-utp/Hybrid-Verification/blob/7baedd092dff0182336ef0bb6251fd8beff6a1cc/HS_Lens_Spartan.thy\#L456}
\begin{equation} \label{eq:frame-rule}
  \prog \nmods A \land (- A) \unrest I \land \hoaretriple{P}{\prog}{Q} \implies \hoaretriple{P \land I}{\prog}{Q \land I}
\end{equation}

\vspace{-\baselineskip}
\subsubsection{Projections.}

Reasoning with frames often requires localising variables to part of the
store. In particular, in \S\ref{sec:dyn-sys} we partition the store into
continuous and discrete parts, and localise continuous variables to the former
to describe their derivatives. Formally, we may use a frame lens
$A : \mathcal{C} \lto \src$ from the global store $\src$ onto a local store
$\mathcal{C}$. Local reasoning within $A$ requires \emph{lens
  quotient}~\cite{Foster2020-LocalVars}, $\lens \lquot A$, which localises a lens
$\lens : T \lto \src$ to a lens $T \lto \csrc$. Assuming $\lens \preceq A$, it obtains
$\lens_1 : T \lto \csrc$ such that $\lens = \lens_1 \lcomp A$. For example,
$v_x \lquot \vec{v} = \Pi(1)$ with $\csrc = \real^n$. Quotient cannot be defined
by, e.g. $\lens \lcomp B$ for $B : \src \lto \csrc$, since
$\csrc$ is smaller than $\src$, and so $B$ cannot satisfy the lens axioms. \hfill\isalink{https://github.com/isabelle-utp/Optics/blob/6e24cde61989a79f7601acc537dd2ee9fdf3f4f6/Lens_Algebra.thy\#L83}

%% file: sec/prog-exp.tex
Next we present our new expression component for Isabelle that
supports seamless transformations between the intuitive expression
syntax often used in verification approaches and the functions
$\src\rightarrow \view$ mimicking them in our shallow embedding. Naive
uses of such functions may pollute specifications with
$\lambda$-binders, e.g. requiring
$\lambda \skey{s}.\, \lget_x~\skey{s} + \lget_y~\skey{s}$ instead of
$x + y$. Isabelle's syntax translations allow relating syntactic and
semantic representations, and ultimately designing interfaces to
modelling languages such as Modelica and Matlab.

\vspace{-\baselineskip}
\subsubsection{Syntax translation.}

Isabelle implements a multi-stage syntax pipeline. Unicode strings are parsed
and transformed into ``pre-terms''~\cite{Kuncar2019-IsabelleFoundation}:
elements of the ML \texttt{term} type containing syntactic constants. These must
be mapped to previously defined semantic constants by syntax translations,
before they can be checked and certified in the Isabelle kernel. Printing
reverses this pipeline, mapping terms to strings. The pipeline supports a host
of syntactic constructions.

We reuse this pipeline with small modifications for our bidirectional expression
transformation, including pretty printing. We subject pre-terms to a lifting
process, which replaces free variables and constants, and inserts store
variables ($\skey{s}$) and $\lambda$-binders. Its implementation uses the
syntactic annotation $\ebrack{t}$ to lift the syntactic term $t$ to a semantic
expression in the syntax translation rules 
$$\begin{array}{ccc}
  \ebrack{t} \syneq [\ebracku{t}]_{\skey{e}},  &\quad \ebracku{x} \syneq
          \begin{cases}
            \lambda \skey{s}.\, \lget_x~\skey{s} & \text{if } x \text{ is a lens} \\
            \lambda \skey{s}.\, x & \text{otherwise}
          \end{cases} &\quad \ebracku{f~t} \syneq \lambda \skey{s}.\, f~(\ebracku{t}~\skey{s}),
  \end{array}$$
  where $p \syneq q$ means that pre-term $p$ is translated to term $q$, and $q$
  printed as $p$. Moreover, $[-]_{\skey{e}}$ is a constant that marks lifted
  expressions that are embedded in terms. The pretty printer can then recognise
  a lifted term and print it. \hfill\isalink{https://github.com/isabelle-utp/Shallow-Expressions/blob/bddbf31a67859011c81e16ad3d6723d66ed9591e/Expressions.thy\#L58}

  Intuitively, $\ebrack{t}$ is processed as follows. The syntax processor first parses a
  pre-term from string $t$. Then our parse translation traverses its
  syntax tree. Whenever it encounters a free variable $x$, the type
  system determines from the context whether it is a lens, in which
  case a $\lget_x$ is inserted. Otherwise it is left unchanged as a
  logical variable. Function applications are left unchanged by
  $\rightleftharpoons$, except for expression constructs like
  $e[v/x]$.  For program variables $x$ and $y$ and logical variable
  $z$, e.g.,
  $\ebrack{(x + y)^2 / z} \rightleftharpoons [\lambda \skey{s}.\,
  (\lget_x~\skey{s} + \lget_y~\skey{s})^2 / z]_{\skey{e}}$. Once an
  expression has been processed, the resulting $\lambda$-term is
  enclosed in $[-]_{\skey{e}}$.

  In assignments $x := e$ and substitutions $[x \smapsto e]$, $e$ is
  lifted transparently without user annotations. We can also lift correctness
  specifications and allow intuitive parsing of assertions:
  $\hoaretriple{P}{\prog}{Q} \syneq \ebrack{P} \le \wlp\, \prog\, \ebrack{Q}$. \hfill\isalink{https://github.com/isabelle-utp/Hybrid-Verification/blob/7baedd092dff0182336ef0bb6251fd8beff6a1cc/HS_Lens_Spartan.thy\#L67}

\vspace{-\baselineskip}
\subsubsection{Substitution.}  

Though our expressions are functions, we can mimic syntactic substitution using
the rule
$\ebrack{\lambda \skey{s}.\, e(\skey{s})}[v/x] = \ebrack{\lambda \skey{s}.\,
  e(\lput_x~(v~s)~\skey{s})}$. This pushes the substitution through the
expression marker, by applying an update to the store. This rule preserves the
lifted syntax, e.g.  $((x + y)^2 / c)_e [2 \cdot x / y] $
simplifies to $((3 \cdot x)^2 / c)_e$. We can also use substitutions to
determine whether a variable is used:
$(x \unrest e) \Leftrightarrow \forall v.\, (e[v/x]) = e$: $e$ does not depend
on $x$ if substituting any value $v$ for it leaves $e$
(\S\ref{sec:hyb-store}). For example, $\ebrack{5}[v/x] = \ebrack{5}$, and thus
$x\unrest 5$. Checking $\unrest$ is usually automatic. More generally, lenses
provide enough structure for simulating many standard syntactic
manipulations semantically to support VCG. \hfill\isalink{https://github.com/isabelle-utp/Shallow-Expressions/blob/bddbf31a67859011c81e16ad3d6723d66ed9591e/Substitutions.thy\#L247}

%% file: sec/diff-eq.tex
We extend our components for the continuous dynamics with a notion of
\emph{function framing} that projects to parts of the store as
outlined in \S\ref{sec:hyb-store}. This supports local reasoning where
evolution commands modify only continuous variables and leave discrete
ones---outside the frame---unchanged. We also introduce \emph{framed
  differentiation} to calculate differential invariants equationally,
and derive a ghost rule~\cite{Platzer18} that expands ODEs with fresh
variables.

\vspace{-\baselineskip}

\subsubsection{Framed vector fields.}

We fix the continuous part $\csrc$ of hybrid store $\src$ with suitable
topological structure for continuous variables and vector fields
(e.g. $\real^n$) and supply a lens
$\lens : \csrc \lto \src =
(\lput_\lens:\csrc\rightarrow\src\rightarrow\src,\lget_\lens:\src\rightarrow\csrc)$. We
can compose this data with any $f : \src \rightarrow \src$---\emph{frame} it
with $\lambda$---to $f_\lambda : \src \rightarrow \csrc \rightarrow \csrc$ such
that
$f_\lambda\, s = \lget_\lambda\circ f\circ (\lambda v.\
\lput_\lambda\,v~s)$. For $\csrc\subseteq\src$, $f_\lens\, s$ is thus the
restriction of $f$ to $\csrc$ supplied with the full store $s$ before
restriction.  For example for $\src = \real^2 \times \real^2\times \src'$,
$\lambda: \real^2\times \real^2 \lto \src = (\vec{p} \lplus \vec{v})$ frames
$f:\src\to\src = [(\vec{p}, \vec{v}) \smapsto (\vec{v}, 0)]$ which behaves 
as the identity function on $\src'$, and thus
$f_\lambda\, s: \real^2 \times \real^2 \rightarrow \real^2 \times \real^2$.
Similarly, if $f:T \rightarrow \src\rightarrow\src$ is a vector field
and $s\in\src$, then
$\lambda t.\ (f\, t)_\lens\, s: T \rightarrow \csrc\rightarrow\csrc$
is the \emph{framed vector field} $f_\lens$ for $f$ and $s$. We can
supply $f$ as a substitution
$\lambda t.\, [\vec{x}\smapsto \vec{e}\, t]$ that describes the ODEs
$\vec{x}'\, t=\vec{e}\, t$ quite naturally, after framing. \hfill\isalink{https://github.com/isabelle-utp/Hybrid-Verification/blob/7baedd092dff0182336ef0bb6251fd8beff6a1cc/HS_Lens_ODEs.thy\#L17}

\vspace{-\baselineskip}
\subsubsection{Specifying evolution commands.}

We have formalised the generalised orbits in the semantics for
$x' = f \, \&\, G$ in \S\ref{sec:prelim} as
$\skey{g-orbital}$~\cite{MuniveS19}. Here we add a framed version,
$\skey{g-orbital-on}\,\lambda~f\,G~U\,S\,t_0:\src\to \power\, \src$
defined as the image under $(\lambda v.\ \lput_\lambda\, v~s)$ of
$\skey{g-orbital}$ applied to $f_\lens$,
$(\lambda v.\ G\, (\lput_\lens v\, s))$, $U$, $S$, $t_0$, and
$(\lget_\lens\, s)$, where $S\subseteq\csrc$ is the codomain of
$f_\lens$.  The application of $\skey{g-orbital}$ to $f_\lens$ makes
it a state transformer on $\csrc$, while its image under
$(\lambda v.\ \lput_\lambda\, v~s)$ lifts it back to $\src$.  VCG with
the first workflow and $\skey{g-orbital-on}$ then remains as outlined
in \S\ref{sec:intro}. Users need to supply flows and constants for
Lipschitz continuity in order to obtain $\wlp$'s as in
\S\ref{sec:prelim}.  We provide tactics that may automate this process
in \S\ref{sec:proof}. \hfill\isalink{https://github.com/isabelle-utp/Hybrid-Verification/blob/7baedd092dff0182336ef0bb6251fd8beff6a1cc/HS_Lens_ODEs.thy\#L30}

With Isabelle's syntax translations, we can specify
\skey{g-orbital-on} naturally as
$\{x_1' = e_1, x_2' = e_2, \cdots, x_n' = e_n ~|~ G
\mathop{\,\skey{on}\,} U\,S\, \mathop{\,@\,} t_0\},$ where each $x_i$
is a summand of the lens $\{x_1, \cdots, x_n\}$. We use this notation in
all our examples of \S\ref{sec:case-studies}.  Users can thus declare 
the ODEs in evolution commands coordinate-wise with lifted expressions 
$e_i$ ranging over $\mathcal{S}$, and the other parameters $G$, $U$, 
$S$ and $t_0$ which can be omitted. Their omission defaults them to 
constantly true, $\{t.\ t\geq 0\}$, $\csrc$ and $0$ respectively. We can 
often collapse further to $(x_1', \cdots, x_n') = (e_1, \cdots, e_n)$ and 
obtain the following framing result:
$\{\vec{x}' = \vec{e} ~|~ G \mathop{\,\skey{on}\,} U\,S\,
\mathop{\,@\,} t_0\} \nmods (- \vec{x})$, as variables outside of
$\vec{x}$ do not change during evolution and hence, by~\eqref{eq:frame-rule} 
in \S\ref{sec:hyb-store}, any $I$ specified only over discrete variables 
is an invariant (see~\eqref{law:dD} below). We can also specify evolutions using
flows with the following notation
$\{\texttt{EVOL}\, (x_1, \cdots, x_n) = (e_1\, \tau, \cdots, e_n\,
\tau) \,|\, G\}$, which also carries a frame. \hfill\isalink{https://github.com/isabelle-utp/Hybrid-Verification/blob/7baedd092dff0182336ef0bb6251fd8beff6a1cc/HS_Lens_Spartan.thy\#L533}

\vspace{-\baselineskip}

\subsubsection{Framed derivatives.}

Isabelle supports Fr\'echet derivatives and we are localising them by
framing. These derivatives are defined for functions between normed
vector spaces or Banach spaces. Recall that, for a Fr\'echet
differentiable function $F:\csrc\to \mathcal{T}$ at $s\in\csrc$, with
$\csrc\subseteq \src$ open and $\src,\mathcal{T}$ Banach spaces, its derivative
$\fderiv\, F\, s$ is a continuous linear operator in
$\csrc\to \mathcal{T}$~\cite{Cheney01}. If $F$ is differentiable everywhere,
$\fderiv\, F:\csrc\to L(\csrc, \mathcal{T})$, where $L(\csrc, \mathcal{T})$ is the
subspace of continuous linear operators in $\csrc\to \mathcal{T}$. For
finite-dimensional spaces, $\fderiv\, F\, s$ is the Jacobian of $F$ at
$a$; compositions with unit vectors yield partial derivatives and the
sum over these along the coordinates of a vector yield directional
derivatives.

Fix $e : \src \rightarrow \mathcal{T}$ with restriction
$e{|^s_A}: \csrc\to \mathcal{T} = e \circ (\lambda v.\ \lput_A~v~s)$
differentiable everywhere, variable set lens $A$ and function
$f:\src\rightarrow\src$. The \emph{Fr\'echet derivative}
$\lderiv{f}{e}{A}:\src \rightarrow \mathcal{T}$ \emph{of} $e$ \emph{at} $s$
\emph{framed by} $A$ \emph{in direction} $f$ is then defined as \hfill\isalink{https://github.com/isabelle-utp/Hybrid-Verification/blob/7baedd092dff0182336ef0bb6251fd8beff6a1cc/HS_Lie_Derivatives.thy\#L107}
\begin{equation*}
 \lderiv{f}{e}{A}\, s =  \fderiv~{e{|^s_A}}\, (\lget_A~s)\, (\lget_A~(f~s)).
\end{equation*}
Here, $\lget_A~(f~s)$ is a vector in $\csrc$. Intuitively, in the
finite dimensional case, $\fderiv~{e{|^s_A}}\, (\lget_A~s):\csrc \to \src$
corresponds to a Jacobian and $\lget_A~(f~s)$ the vector
associated by $f$ to $s$ in $\csrc$ along which the directional
derivative is taken. From the user perspective, after framing, $f$ supplies 
the vector field $f_A$ representing the ODEs, $s$ the 
values of the discrete variables, and $e$, the expression to
differentiate.

\vspace{-\baselineskip}

\subsubsection{Computing framed derivatives.}
We can calculate framed derivatives equationally. For lenses
$x : V \lto \mathcal{S}$, $A : \mathcal{C} \lto \src$, function
$f:\src\rightarrow\src$, and expressions
$k, e_1, e_2 : \mathcal{S} \rightarrow T$,
which when framed by $A$ become everywhere differentiable, \hfill\isalink{https://github.com/isabelle-utp/Hybrid-Verification/blob/7baedd092dff0182336ef0bb6251fd8beff6a1cc/HS_Lie_Derivatives.thy\#L132}
  \begin{align}
    \lderiv{f}{k}{A} &= 0 & \text{if } A \unrest k \label{eq:const} \\
    \lderiv{f}{x}{A} &= 0 &\text{if } x \lindep A \label{eq:disc-var} \\
    \lderiv{f}{x}{A} &= \substlk{f}~x & \text{if } x \preceq A \text { and } x \lquot A \text{ is bounded linear} \label{eq:cont-var} \\
    \lderiv{f}{e_1 + e_2}{A} & \omit\rlap{$\hspace{.65ex}= (\lderiv{f}{e_1}{A}) + (\lderiv{f}{e_2}{A})$}\label{eq:plus} \\
    \lderiv{f}{e_1 \cdot e_2}{A} &\omit\rlap{$\hspace{.65ex}= (e_1 \cdot \lderiv{f}{e_2}{A}) + (\lderiv{f}{e_1}{A} \cdot e_2)$} \\
    \lderiv{f}{e^n}{A} &\omit\rlap{$\hspace{.65ex}=n \cdot (\lderiv{f}{e}{A}) \cdot e^{(n-1)}$} \label{eq:power} \\
    \lderiv{f}{\ln(e)}{A} &= (\lderiv{f}{e}{A}) / e & \text{if } e > 0 \label{eq:ln}
 \end{align}
Laws (\ref{eq:const}) and (\ref{eq:plus}-\ref{eq:ln}) are framed analogues of
known derivative rules. Laws (\ref{eq:disc-var}) and (\ref{eq:cont-var}) state
that the derivative of a discrete variable is zero, while that for a continuous
variable is extracted from $f$ as explained in \S\ref{sec:hyb-store}. For the
latter, we need to show that $x \preceq A$ and that $x$ localised to $\csrc$ 
by $A$ ($x \lquot A$) is a linear lens, that is,
$\lget_{x\lquot A}$ is a continuous linear operator, implying that the
expression $(x)_e$ is Fr\'{e}chet differentiable.
With these laws calculations are equational 
$\lderiv{[x \smapsto 1]}{x^2}{x} = 2\cdot(\lderiv{[x \smapsto 1]}{x}{x})\cdot x=
  2\cdot 1\cdot x=2x$. Observe that changing the lens changes 
  the part of $\src$ described by $\csrc$ and thus, the value of the
  framed derivative: with lens $y\lindep x$ we get $\lderiv{[x \smapsto 1]}{x^2}{y} = 2\cdot(\lderiv{[x \smapsto 1]}{x}{y})\cdot x= 2\cdot 0\cdot x=0$. Also, changing the directionality changes the derivative:
  $\lderiv{[x \smapsto 2]}{x^2}{x} = 2\cdot(\lderiv{[x \smapsto 2]}{x}{x})\cdot x=
  2\cdot 2\cdot x=4x$. \hfill\isalink{https://github.com/isabelle-utp/Hybrid-Verification/blob/7baedd092dff0182336ef0bb6251fd8beff6a1cc/HS_Lens_Spartan_Ex.thy\#L84}

\vspace{-\baselineskip}
\subsubsection{Invariance checking.}

The fact that $I$ is an invariant for vector field $f$, as outlined in
\S\ref{sec:intro}, has so far been modelled as $\textit{diff-inv}$.
We now supply a framed variant
$\textit{diff-inv-on}\,I\,\lambda\,f\,U\,S\,t_0\,G$, replacing
$\skey{g-orbital}$ with
$\skey{g-orbital-on}$ in the definition of invariance, where again,
$\lambda: \mathcal{C} \lto \src$ projects onto
$\mathcal{C}$ and the others parameters also remain unchanged. The
adjunction between backward diamonds and
$\wlp$'s still translates differential invariance to Hoare triples as
in the discussion of invariant sets of \S\ref{sec:prelim}: \hfill\isalink{https://github.com/isabelle-utp/Hybrid-Verification/blob/7baedd092dff0182336ef0bb6251fd8beff6a1cc/HS_Lens_Spartan.thy\#L757}
\begin{equation}\label{thm:loc-dinv}
  \{I\}\, \{\vec{x}' = e ~|~ G \mathop{\,\skey{on}\,} U\,S\,
  \mathop{\,@\,} t_0\}\, \{I\} \Leftrightarrow  \textit{diff-inv-on}\,I\,\vec{x}\,(\lambda y.\, [\vec{x} \leadsto e])\,U\,S\,t_0\,G.
\end{equation}
It allows us to derive the \dL-style inference rules shown
below. These are all proved as theorems using \textit{diff-inv-on},
along with various rules for weakening and strengthening to support
VCG. \hfill\isalink{https://github.com/isabelle-utp/Hybrid-Verification/blob/7baedd092dff0182336ef0bb6251fd8beff6a1cc/HS_Lens_Spartan.thy\#L960}
  \begin{gather}
    \displaystyle
    \qquad\qquad~~ \frac{
      G \implies \left(\lderivn{[\vec{x} \smapsto e]}{\,e_1}{\vec{x}}\right) \propto^* \left(\lderivn{[\vec{x} \smapsto e]}{\,e_2}{\vec{x}}\right)
    }{
      \hoaretriple{e_1 \propto e_2}{\{\vec{x}' = e \,|\, G\}}{e_1 \propto e_2}
    } ~ \propto\in\{=,\leq,<\} \label{law:dI} \\[1ex]
    \frac{
      \hoaretriple{I_1}{\{\vec{x}' = e \,|\, G\}}{I_1} \quad \hoaretriple{I_2}{\{\vec{x}' = e \,|\, G \land I_1\}}{I_2}
      }{
      \hoaretriple{I_1 \land I_2}{\{\vec{x}' = e \,|\, G \}}{I_1 \land I_2}
      } \label{law:dC} \\[1ex]
      \frac{
      \vec{x} \unrest I \quad \hoaretriple{P}{\{\vec{x}' = e \,|\, G\}}{Q}
    }{
      \hoaretriple{P \land I}{\{\vec{x}' = e \,|\, G\}}{Q \land I}
    } \label{law:dD}  \\[1ex]
    \frac{
      y\!\lindep\!\vec{x} \quad y\unrest (G, e) ~~ \hoaretriple{I}{\{(\vec{x}, y)' = (e, k \cdot y) \,|\, G\}}{I}
      }{
      \hoaretriple{\exists v.\, I[v/y]}{\{\vec{x}' = e \,|\, G\}}{\exists v.\, I[v/y]}
    } \label{law:dG}
  \end{gather}
   Rule~\eqref{law:dI} performs differential induction on (in)equalities~\cite{Platzer12}. There,
  $\propto^*$ is $\leq$ if $\propto$ is $<$, and $\propto$ otherwise: if the
  framed derivatives of two expressions satisfy an (in)equality, then a
  corresponding (in)equality is an invariant. More complex invariants
  (e.g. boolean combinations) reduce to these cases where computations take
  place~\cite{MuniveS19}. If Fr\'echet derivatives are not defined
  everywhere we can still follow the procedure for invariant reasoning
  in~\cite{MuniveS19}. Rule~\eqref{law:dC} is differential cut, framed but
  unchanged from~\cite{MuniveS19} otherwise. It accumulates
  invariants in the guard sequentially during VCG. Framed differential weakening is also
  derived as before~\cite{MuniveS19}, but omitted for space
  reasons. Rule~\eqref{law:dD} is a frame rule that
  discharges invariants if they only refer to discrete variables,
  $\vec{x} \unrest I$. 
  
  Finally, we have also derived~\eqref{law:dG}, a framed variant of
  the differential ghost rule of \dL~\cite{Platzer18}. In \dL, it substantially
 expands the reasoning power with
invariants~\cite{PlatzerT18}. It is used to transform an invariant for
a vector field $f$ into an invariant of $f$ extended with a fresh
variable $y$ and its derivative. The extension makes the new invariant easy to prove
with rules~\eqref{law:dI},~\eqref{law:dC} and~\eqref{law:dD}.
Formally, rule~\eqref{law:dG} says that $I$, with $y$ abstracted, is
an invariant for a system of ODEs with variables in $\vec{x}$ if it is
also an invariant for the same system but with $y$ as a fresh program
variable, $\vec{x} \oplus y$, satisfying $y'=k\cdot y$ for some
constant $k$. Lens $y$ must already exists in $\src$, but we can
always expand the store with new
variables~\cite{Foster2020-LocalVars}. We limit the
derivative of $y$ to be $k\cdot y$, but we will generalise in the future. \hfill\isalink{https://github.com/isabelle-utp/Hybrid-Verification/blob/7baedd092dff0182336ef0bb6251fd8beff6a1cc/HS_Lens_Spartan.thy\#L972}

%% file: sec/proof.tex
We have turned the results from \S\ref{sec:prog-exp} and \S\ref{sec:dyn-sys}
into automated proof methods for hybrid programs using the Eisbach
tool~\cite{MatichukMW16}. These increase proof automation for both our workflows
relative to~\cite{MuniveS19}. Our proof methods use our baseline tactic
$\skey{expr-auto}$, which targets equalities and inequalities for shallow
expressions.

\vspace{-\baselineskip}

\subsubsection{\skey{hoare-wp-auto}.}
First we supply a proof method for automatic structural VCG in our shallow
approach.  To discharge partial correctness specifications
$\hoaretriple{P}{S}{Q}$, it (1) computes $\wlp\, S\, Q$ by
simplification, (2) reduces substitutions and side-conditions, (3)
applies $\skey{expr-auto}$ to the resulting proof goals.  Our set-up
ensures that the expression syntax is not exploded to HOL terms until
step (3), which leads to readable data-level proofs. \hfill\isalink{https://github.com/isabelle-utp/Hybrid-Verification/blob/7baedd092dff0182336ef0bb6251fd8beff6a1cc/HS_Lens_Spartan.thy\#L986}

\vspace{-\baselineskip}

\subsubsection{\skey{dInduct}.}

To prove goals $\hoaretriple{I}{\{\vec{x}' = e ~|~ G \}}{I}$, it (1)
applies Law~\ref{law:dI} to get a framed derivative expression, and (2)
applies derivative calculation laws (Laws~\ref{eq:const}-\ref{eq:ln}),
substitution laws, and basic simplification laws. This yields  derivative-free
arithmetic equalities or inequalities.
For cases requiring deduction, we supply $\skey{dInduct-auto}$, which applies
\skey{expr-auto} after \skey{dInduct}, plus further simplification lemmas from
\textsf{HOL-Analysis}. Ultimately such heuristics should be augmented with
decision procedures~\cite{Holzl2009-Approximate,Blanchette2016Hammers,Li2017-Poly,Cordwell2021BKR}, as
oracles or verified components. \hfill\isalink{https://github.com/isabelle-utp/Hybrid-Verification/blob/7baedd092dff0182336ef0bb6251fd8beff6a1cc/HS_Lie_Derivatives.thy\#L374}

While \skey{dInduct-auto} suffices for simpler examples, differential induction
must often be combined with weakening and cut rules. This leads to another more
versatile proof method, using Isabelle's Eisbach~\cite{MatichukMW16} proof
method language.

\vspace{-\baselineskip}

\subsubsection{\skey{dInduct-mega}.}

The following steps are executed iteratively until all goals are proved or no
rule applies: (1) try any facts labelled with attribute \texttt{facts}, (2) try
differential weakening to prove the goal, (3) try differential cut
(Law~\ref{law:dC}) to split it into two differential invariants, (4) try
\skey{dInduct-auto}. The rules are applied using some backtracking, so that if
one fails, another one is tried. This automatically discharges many differential
invariants, as shown in \S\ref{sec:case-studies}. \hfill\isalink{https://github.com/isabelle-utp/Hybrid-Verification/blob/7baedd092dff0182336ef0bb6251fd8beff6a1cc/HS_Lie_Derivatives.thy\#L381}

\vspace{-\baselineskip}

\subsubsection{\skey{local-flow}.}

While the proof methods so far describe the second workflow using framed
versions of the inference rule of \dL, the first workflow of the
framework~\cite{MuniveS19} supports verification with certified solutions
(flows), which can be supplied using a CAS~\cite{Hickman2021-CASIsabelle}. We
have developed a proof method called \skey{local-flow} for certifying that a
flow is the unique solution to an ODE. As this involves supplying a suitable
Lipschitz constant, we have written a proof method \skey{local-flow-auto} that tries
several such constants, such as $0.5$, $1$ and $2$. The Picard-Lindel\"of
theorem can then be supplied to \skey{hoare-wp-auto} and used to replace any ODE
system within by its local flow. This can lead to simpler VCs than
differential
induction. \isalink{https://github.com/isabelle-utp/Hybrid-Verification/blob/7baedd092dff0182336ef0bb6251fd8beff6a1cc/HS_Lens_Spartan.thy\#L994}

%% file: sec/case-studies.tex
Finally we evidence the benefits of our extensions to the
framework~\cite{MuniveS19} by examples. We illustrate the flexibility
of our store model, the usability provided by shallow expressions, the
local reasoning provided by frames and framed derivatives and the
automation provided by our proof methods.

\vspace{-\baselineskip}
\subsubsection{Circular pendulum.} We begin with a small example
explaining \skey{dInduct}. Consider a circular pendulum in variables
$x, y : \real \lto \src$, and constant $r : \real$ for the radius.  We
use \skey{dInduct} to verify a simple invariant for its vector field. \hfill\isalink{https://github.com/isabelle-utp/Hybrid-Verification/blob/7baedd092dff0182336ef0bb6251fd8beff6a1cc/HS_Lens_Examples.thy\#L69}
{\small
\begin{alltt}
\isakwmaj{lemma} pend:"\{r\(\sp{2}\) = x\(\sp{2}\)+y\(\sp{2}\)\} \{x' = y, y' = -x\} \{r\(\sp{2}\) = x\(\sp{2}\)+y\(\sp{2}\)\}" \isakwmaj{by} dInduct
\end{alltt}
}
\noindent First, the tactic yields
$\lderiv{f}{r^2}{\{x,y\}} = \lderiv{f}{x^2 + y^2}{\{x, y\}}$, where
$f = [x \leadsto y, y \leadsto -x]$ is the vector field. It then
computes $2 \cdot y \cdot x + 2 \cdot (-x ) \cdot y = 0$ using
differentiation and substitution rules, which is discharged
automatically by arithmetic reasoning~\cite{MuniveS19}.

\vspace{-\baselineskip}
\subsubsection{Water tank.}

Next, we formalise the classic water tank example, which requires the water
level $h$ to remain between bounds $h_l\leq h\leq h_u$. A controller turns a
water pump on and off to regulate $h$. The dataspace is described below. \hfill\isalink{https://github.com/isabelle-utp/Hybrid-Verification/blob/7baedd092dff0182336ef0bb6251fd8beff6a1cc/HS_Lens_Examples.thy\#L409}
{\small
\begin{alltt}
\isakwmaj{dataspace} water\_tank = \isakwmin{constants} \(h\sb{l}\)::\(\real\) \(h\sb{u}\)::\(\real\) \(c\sb{o}\)::\(\real\) \(c\sb{i}\)::\(\real\)
  \isakwmin{assumes} co:"0 < \(c\sb{o}\)" \isakwmin{and} ci:"\(c\sb{o}\) < \(c\sb{i}\)" \isakwmin{variables} flw::\(\bool\) h::\(\real\) h\(\sb{m}\)::\(\real\) t::\(\real\)
\end{alltt}}
\noindent Constants $c_o$ and $c_i$ indicate rates of outflow and inflow (when
the pump is on). Variable \texttt{flw} is for the water pump, $h_m$ for water
level measurements, and $t$ for time. Previously, every variable had to be a
real number. Now, we provide a clear separation between discrete and
continuous variables; \texttt{flw} is of type $\bool$.

With framed ODEs, we need only assign derivatives to variables $h$ and
$t$, and the other variables are implicitly immutable during
evolution: \hfill\isalink{https://github.com/isabelle-utp/Hybrid-Verification/blob/7baedd092dff0182336ef0bb6251fd8beff6a1cc/HS_Lens_Examples.thy\#L436}
{\small
\begin{alltt}
\isakwmaj{abbreviation} "dyn \(\equiv\) IF flw THEN \{h' = -\(c\sb{o}\), t' = 1 | t \(\leq\) (\(h\sb{l}\)-\(h\sb{m}\))/(-\(c\sb{o}\))\}
                    ELSE \{h' = \(c\sb{i}\)-\(c\sb{o}\), t' = 1 | t \(\leq\) (\(h\sb{u}\)-\(h\sb{0}\))/(\(c\sb{i}\)-\(c\sb{o}\))\}"
\end{alltt}}
\noindent These two ODEs model the dynamics when the inflow is on and off,
respectively. They can be written intuitively due to our expression model
(\S\ref{sec:prog-exp}). The frame is inferred as $\{h, t\}$, since
we only assign derivatives to them, and so $h_m$ and $flw$ remain
unchanged during evolution, as the following theorem confirms: \hfill\isalink{https://github.com/isabelle-utp/Hybrid-Verification/blob/7baedd092dff0182336ef0bb6251fd8beff6a1cc/HS_Lens_Examples.thy\#L441}
{\small
\begin{alltt}
\isakwmaj{lemma} nm: "dyn nmods \{flw, h\(\sb{m}\)\}" \isakwmaj{by} (simp add: closure)
\isakwmaj{lemma} "\{flw = F\} dyn \{flw = F\}" \isakwmaj{by} (rule nmods_invariant[OF nm], unrest)
\end{alltt}
}
\noindent Specifically, \texttt{dyn} modifies neither \texttt{flw} nor $h$, and
so \texttt{flw} keeps its initial value \texttt{F}, which is
recognised as a logical variable. The latter theorem is proved using our frame rule \eqref{eq:frame-rule}.
Next, we can specify our controller: \hfill\isalink{https://github.com/isabelle-utp/Hybrid-Verification/blob/7baedd092dff0182336ef0bb6251fd8beff6a1cc/HS_Lens_Examples.thy\#L432}
{\small
\begin{alltt}
\isakwmaj{abbreviation} "ctrl \(\equiv\) (t,h\(\sb{m}\))::=(0,h); IF \(\neg\)flw \(\land\) h\(\sb{m}\)\(\leq\)H\(\sb{l} +1\) THEN flw::=True
  ELSE IF \(flw \land h\sb{m}\geq h\sb{u}-1\) THEN \(flw\) ::= False ELSE skip)"
\end{alltt}}

\noindent This first assigns $0$ and $h$ to $t$ and $h_m$,
respectively, to reset the time and measure the water level. Then, if
the inflow is off and the height is getting close to the minimum, then
the inflow is enabled. Otherwise, if the level is getting near the
maximum, then it is disabled. If neither is true, then we skip.

Using the second workflow, we use differential induction to discharge
invariants for the tank dynamics. We focus on the invariant
when the inflow is enabled:
{\small
\begin{alltt}
\isakwmaj{lemma} "\{0 \(\le\) t \(\land\) h = (c\(\sb{i}\) - c\(\sb{o}\))*t + h\(\sb{m}\) \(\land\) H\(\sb{l}\) \(\le\) h \(\land\) h \(\le\) H\(\sb{u}\)\}
       \{h` = c\(\sb{i}\) - c\(\sb{o}\), t` = 1 | t \(\le\) (H\(\sb{u}\) - h\(\sb{m}\)) / (c\(\sb{i}\) - c\(\sb{o}\))\}
       \{0 \(\le\) t \(\land\) h = (c\(\sb{i}\) - c\(\sb{o}\))*t + h\(\sb{m}\) \(\land\) H\(\sb{l}\) \(\le\) h \(\land\) h \(\le\) H\(\sb{u}\)\}"
  \isakwmaj{using} ci \isakwmaj{by} dInduct_mega
\end{alltt}}

  \noindent This shows one step of the verification. We prove several
  invariants, using $c_i < c_o$. The key is to prove that
  $h = (c_i - c_o)\cdot t + h_m$, which gives the solution for $h$ and allows us
  to bound how much the water rises. \skey{dInduct-mega} automates the proof
  with successive differential cuts. The system proof is concluded with \skey{hoare-wp-auto} to verify
  the controller. The verification is automated by a high-level proof method
  (\skey{dProve}):
  \hfill\isalink{https://github.com/isabelle-utp/Hybrid-Verification/blob/434303bda3aa79b1c0b6f56de40b84622f6a49a5/HS_Lens_Examples.thy\#L462}
  {\small
\begin{alltt}
\isakwmaj{lemma} tank_correct:
  "\{t = 0 \(\land\) h = h\(\sb{m}\) \(\land\) H\(\sb{l}\) \(\le\) h \(\land\) h \(\le\) H\(\sb{u}\)\} 
    LOOP ctrl ; dyn INV (0\(\le\)t \(\land\) h = ((flw*c\(\sb{i}\))-c\(\sb{o}\))*t + h\(\sb{m}\) \(\land\) H\(\sb{l}\)\(\le\)h \(\land\) h\(\le\)H\(\sb{u}\))
   \{H\(\sb{l}\) \(\le\) h \(\land\) h \(\le\) H\(\sb{u}\)\}" \isakwmaj{using} ci co \isakwmaj{by} dProve
\end{alltt}
}

\noindent We need to supply an extended invariant for both the controller and
dynamics via an annotation. Internally the proof uses the frame rule to
demonstrate that both $\texttt{flw}$ and $\neg \texttt{flw}$ are invariants of
\texttt{dyn}. The earlier lemma is technically not required, as \skey{dProve} itself
invokes \skey{dInduct-mega} during the proof.

We can alternatively verify the controller using the first workflow, with a
solution to the differential equations: \hfill\isalink{https://github.com/isabelle-utp/Hybrid-Verification/blob/434303bda3aa79b1c0b6f56de40b84622f6a49a5/HS_Lens_Examples.thy\#L429}

{\small
  \begin{alltt}
\isakwmaj{lemma} lf:"local_flow_on [h\(\leadsto\)k,t\(\leadsto\)1] (h\(\lplus\)t) UNIV UNIV (\(\lambda\tau\).[h\(\leadsto\)k*\(\tau\),t\(\leadsto\)\(\tau\)+t])"
  \isakwmaj{by} local_flow_auto

\isakwmaj{lemma} "\{h\(\sb{m}\)\(\leq\)h \(\land\) h\(\leq\)h\(\sb{M}\)\} LOOP ctrl;dyn INV (h\(\sb{m}\)\(\leq\)h\(\land\)h\(\leq\)h\(\sb{M}\)) \{h\(\sb{m}\)\(\leq\)h \(\land\) h\(\leq\)h\(\sb{M}\)\}"
  \isakwmaj{using} tank_arith[OF _ co ci] \isakwmaj{by} (hoare_wp_auto local_flow: lf)
\end{alltt}}

\noindent We need to certify the unique (framed) solution for the
water tank vector field using \skey{local-flow-auto}. The loop
invariant need not refer to \texttt{flw} or $h_m$, since these are
discrete. The proof uses \skey{hoare-wp-auto}, which is given the
local flow proof \texttt{lf}, and so can internally replace the ODE
with the flow.

\vspace{-\baselineskip}
\subsubsection{Exponential decay.}

This example contrasts the use of the differential ghost law and solutions in
ODE proof. We wish to show that $x > 0$ is an invariant of $x' = - x$, which is
not immediately obvious because the derivative is negative.

The proof using differential ghost is as follows: $x > 0$ is equivalent to
$xy^2=1$ for some value of $y$. To show that this new property is an invariant
for the evolution command, we use the ghost rule to expand the system of ODEs
into $x'=-x,y'=y/2$, which retains the behaviour of $x$. Then, by standard
reasoning~\cite{MuniveS19}, $xy^2=1$ is an invariant because
$(xy^2)'=x'y^2 + 2xyy' = - xy^2 + xy^2 = 0 = 1'$. The Isabelle proof requires
some interaction: \hfill\isalink{https://github.com/isabelle-utp/Hybrid-Verification/blob/434303bda3aa79b1c0b6f56de40b84622f6a49a5/HS_Exponential_Ex.thy\#L62}
{\small
\begin{alltt}
\isakwmaj{lemma} dG\_example: "\{x > 0\} \{x' = -x\} \{x > 0\}"
  \isakwprf{apply} (dGhost "y" "(\(x*y^2\) = 1)\(\sb{e}\)" "1/2", expr_auto add: exp_arith)
  \isakwprf{apply} (dInduct_auto, simp add: power2_eq_square) \isakwprf{done}
\end{alltt}}
\noindent The first line applies differential ghost, with the fresh
variable $y$, the property $xy^2=1$, and the factor $1/2$ of the ODE
$y'=y/2$. This yields $0 < x \iff \exists v.\, xv^2 = 1$, which we
prove using a lemma (\texttt{exp\_arith}). The last line applies
\skey{dInduct-auto} and a \skey{sledgehammer}~\cite{Blanchette2016Hammers} supplied proof for the
goal $y y \neq y^2 \implies y = 0$.

For the first workflow, $x > 0$ is an invariant because the solution,
$x\, t = x_0 e^{-t}$, is a positive exponential function for $x_0 > 0$. The
proof of invariance follows immediately from this fact and automatically with
our tactic. \hfill\isalink{https://github.com/isabelle-utp/Hybrid-Verification/blob/434303bda3aa79b1c0b6f56de40b84622f6a49a5/HS_Exponential_Ex.thy\#L92}
{\small
\begin{alltt}
\isakwmaj{lemma} flow\_ex:"\{x > 0\} \{x' = -x\}\{x > 0\}" \isakwmaj{by} (hoare\_wp\_auto local\_flow: lf)
\end{alltt}}
\noindent We again supply a certified solution using the theorem
\texttt{lf} (omitted). Alternatively, if users wish to skip the
certification and write the flow directly in the specification, this
is also possible and automatic using notation introduced in\S\ref{sec:dyn-sys}.  
{\small
\begin{alltt}
\isakwmaj{lemma} "\{x > 0\} \{EVOL x = x * exp (- \(\tau\))\} \{x > 0\}" \isakwmaj{by} hoare_wp_auto
\end{alltt}}
 
\vspace{-\baselineskip}
\subsubsection{Autonomous boat.}

The final example~\cite{FosterGC20} is a controller for an autonomous boat using
the second workflow. We model the dynamics and several controller components,
which demonstrates our flexible hybrid store model and local reasoning with our
framework. The boat's objective is to navigate along several way-points, while
avoiding obstacles. It is manoeuvrable in $\real^2$ and has a
rotatable thruster generating a positive propulsive force $\vec{f}$ with maximum
$f_{max}$.  The boat's state is determined by its position $\vec{p}$, velocity
$\vec{v}$, and acceleration $\vec{a}$: \hfill\isalink{https://github.com/isabelle-utp/Hybrid-Verification/blob/03a4a9ddc361bdd3e9c6dd1611920cd9f9fa4773/Boat.thy\#L183}
{\small
\begin{alltt}
\isakwmaj{dataspace} AMV = \isakwmin{constants} S::\(\real\) f\(\sb{max}\)::\(\real\) \isakwmin{assumes} fmax:"f\(\sb{max}\) \(\ge\) 0"
  \isakwmin{variables} p::"\(\real\) vec[2]" v::"\(\real\) vec[2]" a::"\(\real\) vec[2]" \(\phi\)::\(\real\) s::\(\real\)
    wps::"(\(\real\) vec[2]) list" org::"(\(\real\) vec[2]) set" rs::\(\real\) rh::\(\real\)
\end{alltt}
}
\noindent This complex store model consists of a combination of discrete and
continuous variables. Here, $\real~\texttt{vec[n]}$ is a vector of dimension
$n$. In the \isakwmaj{dataspace}, we have a variable for linear speed $s$, and
constant $S$ is the maximum speed. We
also model discrete variables for the way-point path (\texttt{wps}), the
obstacle register (\texttt{org}), and requested speed and heading (\texttt{rs}
and \texttt{rh}).

With the following axiom, we relate $s$, $\phi$ with $\vec{v}$ and constrain $s$. \hfill\isalink{https://github.com/isabelle-utp/Hybrid-Verification/blob/03a4a9ddc361bdd3e9c6dd1611920cd9f9fa4773/Boat.thy\#L205}
{\small 
\begin{alltt}
\isakwmaj{abbreviation} "ax \(\equiv\) (s *\(\sb{R}\) [[sin(\(\phi\)), cos(\(\phi\))]] = v \(\land\) 0 \(\le\) s \(\land\) s \(\le\) S)\(\sp{e}\)"
\end{alltt}}

\noindent This states that \vec{v} is equal to $s$ multiplied by the
heading unit-vector using scalar multiplication (\texttt{*\(\sb{R}\)}) and our
vector syntax, and $0 \le s \le S$. The kinematics
$[\vec{p}',\vec{v}',\vec{a}']^T$ is specified below: \hfill\isalink{https://github.com/isabelle-utp/Hybrid-Verification/blob/03a4a9ddc361bdd3e9c6dd1611920cd9f9fa4773/Boat.thy\#L209}
{\small
\begin{alltt}
\isakwmaj{abbreviation} "ODE \(\equiv\) \{ p` = v, v` = a, a` = 0, \(\phi\)` = \(\omega\), 
                       s` = if s \(\neq\) 0 then (v \(\cdot\) a) / s else \(\lVert\)a\(\rVert\) | @ax \}"
\end{alltt}
}
\noindent We also specify derivatives for $\phi$ and $s$. The former, $\omega$
is the angular velocity, which has the value
$\textit{arccos}((\vec{v} + \vec{a})\cdot \vec{v} / (\lVert \vec{v} + \vec{a}
\rVert \cdot \lVert \vec{v} \rVert))$ when $\lVert \vec{v} \rVert \neq 0$ and
$0$ otherwise. The linear acceleration ($s'$) is calculated using the inner
product of $\vec{v}$ and $\vec{a}$. If the current speed is $0$, then $s'$ is
simply $\lVert a \rVert$. All other variables in the store are implicitly
outside the evolution frame: \hfill\isalink{https://github.com/isabelle-utp/Hybrid-Verification/blob/03a4a9ddc361bdd3e9c6dd1611920cd9f9fa4773/Boat.thy\#L221}

{\small
\begin{alltt}
\isakwmaj{lemma} "ODE nmods \{rs, rh, wps, org\}" \isakwmaj{by} (simp add: closure)
\end{alltt}
}

\noindent The controller for the AMV consists of three parts: \texttt{Navigation} for
way-point following, \texttt{AP} the autopilot proportional controller, and
\texttt{LRE} the safety controller. \texttt{Navigation} and \texttt{LRE} both
supply requested headings and speeds to the \texttt{Autopilot}, which calculates
an acceleration vector for the \texttt{ODE}. For reasons of space, we omit
further details.  We prove some differential invariants of \texttt{ODE}: \hfill\isalink{https://github.com/isabelle-utp/Hybrid-Verification/blob/03a4a9ddc361bdd3e9c6dd1611920cd9f9fa4773/Boat.thy\#L226}
{\small
\begin{alltt}
\isakwmaj{lemma} "\{s\(\sp{2}\) = v \(\cdot\) v\} ODE \{s\(\sp{2}\) = v \(\cdot\) v\}" \isakwmaj{by} (dWeaken, metis orient_vec_mag_n)
\isakwmaj{lemma} "\{a = 0 \(\land\) v = V\} ODE \{a = 0 \(\land\) v = V\}" \isakwmaj{by} (dInduct_mega)
\isakwmaj{lemma} "\{(a = 0 \(\land\) s > 0) \(\land\) \(\phi\) = X\} ODE \{\(\phi\) = X\}" \isakwmaj{by} (dInduct_mega)
\end{alltt}
}
\noindent The first shows that $s^2$ is equal to the inner product of $v$ with
itself, and is charged by differential weakening and a vector lemma. The second
shows that if the acceleration vector is 0, the velocity is not changing. The
third shows, similarly, that the orientation is not changing. These demonstrate
our ability to form differential invariants over vectors, as well as
scalars. We also reason about vectors algebraically without
coordinate-wise decomposition. Since $s$ and $v$ are modified only by the ODE, we
can also show that the first invariant is also an invariant of each other
component using our frame rule; e.g.
$\hoaretriple{s^2 = v}{\texttt{Autopilot}}{s^2 = v}$. Moreover, this is also a
system invariant. This kind of local reasoning makes proof about such
component-based systems tractable.  We check a final \texttt{ODE} invariant:
{\small
\begin{alltt}
\isakwmaj{lemma} "\{a\(\cdot\)v\(\ge\)0 \(\land\) (a\(\cdot\)v)\(\sp{2}\) = (a\(\cdot\)a)\(\cdot\)(v\(\cdot\)v)\} ODE \{a\(\cdot\)v\(\ge\)0 \(\land\) (a\(\cdot\)v)\(\sp{2}\) = (a\(\cdot\)a)\(\cdot\)(v\(\cdot\)v)\}"
  \isakwmaj{by} (dInduct_mega, metis inner_commute)
\end{alltt}
}
\noindent This property tells us that if \vec{v} and \vec{a} have the same
direction, they will continue to do so. The boat may be linearly accelerating or
decelerating, but is not turning. Method \skey{dInduct-mega} produces a proof
obligation relating to inner product, which is discharged with
\skey{sledgehammer}~\cite{Blanchette2016Hammers}. \hfill\isalink{https://github.com/isabelle-utp/Hybrid-Verification/blob/03a4a9ddc361bdd3e9c6dd1611920cd9f9fa4773/Boat.thy\#L233}

%% file: sec/concl.tex
We have transformed an Isabelle framework for the verification of hybrid systems
into a more user-friendly and effective formal method, using Isabelle's syntax
translation mechanisms to interface with more natural modelling and
specification languages for hybrid programs. We have also shown how local
reasoning about hybrid stores can be achieved using lenses, how this leads to
better and more automatic tactics and to more accessible verification
conditions.

Deductive reasoning about hybrid systems is not new, and there is
substantial work supporting this activity through domain-specific
decision procedures.  With PVS, a formalisation for \dL-style
verification by formalising semi-algebraic sets and real analytic
functions is in its early steps~\cite{SlagelWD21}. With Coq, the
ROSCoq framework~\cite{AnandK15} uses Coq's CoRN library of
constructive real numbers to formalise a Logic of Events (LoE) to
reason about hybrid systems. The VeriDrone
project~\cite{RickettsMAGL15} uses the Coquelicot library for a
variant of the temporal logic of actions. Both semantics are very
different from our \dL-inspired one. Yet we view the framework rather
as complementary to \dL's KeYmaera X tool, which brings the benefit of
certified decisions procedures for real arithmetic and a user-friendly
interface. With Isabelle, a term checker for \dL has been formalised
as a deep embedding~\cite{BohrerRVVP17,afp:dgl}, but without aiming at
verification components. A hybrid Hoare logic~\cite{WangZZ15} for
verifying hybrid CSP processes~\cite{LiuLQZZZZ10} has been formalised
as a shallow embedding, but with a very different semantics to our own
Hoare logic~\cite{FosterMS20}.

Work remains to be done for transforming this framework further into
an industrial-strength formal method. This is supported by its
openness and compositionality. Our new hybrid store and extant ODE
components for Isabelle could, for instance, be integrated into the
state transformer semantics~\cite{MuniveS19} with little effort. Our
new expression language could easily be replaced by Modelica
syntax. In the future, one could use Isabelle's code generator to
provide pathways from hybrid programs to verified controller
implementations. Openness implies in particular that anyone interested
in hybrid systems verification with Isabelle could extend, adapt and
contribute to this framework. Its power is only limited by Isabelle's
higher order logic and the mathematical and verification components
that the community provides for it.